\documentclass[12pt]{article}

\def\ii{\'{\i }}

\usepackage{epsfig}

\def\maiorsim{\smash{\mathop{>}\limits_{\raise3pt\hbox{$\sim$}}}}

\def\infinito{\smash{\mathop{\longrightarrow}\limits_{\raise3pt\hbox{$_{y_b \to \infty}$}}}}

\def\doisdeltas{\smash{\mathop{\longrightarrow}\limits_{\raise3pt\hbox{$_{\Delta >\Delta_0 (y_b)}$}}}}

\begin{document}

\title{Limiting fragmentation in heavy-ion collisions and percolation of strings}%
\author{P. Brogueira\footnote{ Departamento de F\ii sica, IST, Av. Rovisco Pais, 1049-001 Lisboa, Portugal}\ , J. Dias de Deus\footnote{CENTRA, Departamento de F\ii sica, IST, Av. Rovisco Pais, 1049-001 Lisboa, Portugal} \ and C. Pajares\footnote{IGFAE and Departamento de Fisica de Particulas, Univ. of Santigo de Compostela, 15706, Santiago de Compostela, Spain}
}

\maketitle

\begin{abstract}
The observed limiting fragmentation of charged particle distributions in heavy ion collisions is difficult to explain as it does not apply to the proton spectrum itself. On the other hand, string percolation provides a mechanism to regenerate fast particles, eventually compensating the rapidity shift (energy loss) of the nucleons. However a delicate energy-momentum compensation is required, and in our framework we see no reason for limiting fragmentation to be exact. A prediction, based on percolation arguments, is given for the charged particle density in the full rapidity interval at LHC energy $(\sqrt s =5500\ GeV)$.
\end{abstract}

\bigskip
\bigskip

Recentely, the phenomenon of limiting fragmentation, [1], or longitudinal scaling, was rediscovered in the framework of high-energy heavy ion collisions [2]. In general the inclusive particle distribution, $dn/dy$, is a function of the central rapidity $y$ and of the centre of mass energy $\sqrt{s}$, or of $\Delta \equiv y-y_b$ and $y_b$, where $y_b$ is the beam rapidity, with $\sqrt{s}=m_b e^{y_b}$. Limiting fragmentation essentially means that as $\Delta$ becomes larger than some $y_b$ dependent threshold $\Delta_0$, $\Delta \geq \Delta_0 (y_b)$, $dn/dy$ becomes a function only of $\Delta$,
$$
{dn\over dy} (\Delta ,y_b) \doisdeltas f (\Delta) \  ,\eqno(1)
$$
independent of $y_b$. As $\Delta_0 (y_b)$ decreases with $y_b$ the region in $\Delta$ of limiting fragmentation increases with the energy.

It has been argued that limiting fragmentation reflects the fast parton distribution in the beam [3]. However, as a sizeable fraction of the fast particles building up the limiting fragmentation behaviour are nucleons (protons), [4], one requires specific parton correlations to generate the observed nucleons. This is done, for instance, in the Dual Parton Model, [5], by introducing valence diquarks which, directly or indirectly, [6], produce baryons, thus preserving the flow of baryon number.

It is important to remark that the proton spectrum at high rapidity does not scale in the sense of limiting fragmentation, [4], presenting a shift $\langle \Delta \rangle_B \equiv \langle y \rangle_B - y_b$ increasing in absolute value with energy, at least for intermediate energies [7]. Theoretically, in QCD evolution models, a transfer of energy and momentum from fast partons to the sea is also expected, [8].

In such circumstances how is it possible to obtain overall limiting fragmentation? A possible solution is given by string percolation models as percolation implies not only a summation in colour, [9,10], but, as well, a summation in momentum, [11,12]. In a sense, percolation is a mechanism for reacceleration of particles. 

In string percolation models strings are produced along the collision axis. In the impact parameter plane if the area of interaction is $\pi R^2$, the projected discs from the strings have an area $\pi r^2$ and there are $\bar N_s$ strings, the transverse density parameter $\eta$,
$$
\eta \equiv ({r\over R})^2 \bar N_s \  , \eqno(2)
$$
is the relevant parameter in percolation. If $\eta \ll 1$ the strings are independent and do not overlap, if $\eta > 1$ the strings fuse and percolate.

When strings overlap, due to random colour summation, the particle rapidity density $dn/dy$ is not the sum of the particle density $\bar n_1$ of each string, and the average transverse momentum $<p_T>$ is not the single string average transverse momentum $\bar p_1$. In general we have
$$
dn/dy =F (\eta) \bar N_s \bar n_1 \ , \eqno(3)
$$
and
$$
<p_T> = \bar p_1 /\sqrt{F(\eta)} \ ,  \eqno(4)
$$
where $F(\eta)$ is the colour summation reduction factor:
$$
F(\eta) \equiv \sqrt{1-e^{-\eta}\over \eta} \ . \eqno(5)
$$
Note that $\eta \to 0$, $F(\eta) \to 1$, and that as $\eta \to \infty$, $F(\eta) \to 1/\eta^{1/2}$.

Regarding the momentum summation, if we have a cluster of $N$ strings, each string made up of partons with Feynman $x$-variable $-x_1$ and $+x_1$, respectively, such that $y_1 =y_b + \ln x_1$, [12], we have for the end (forward rapidity of the $N$-cluster string 
$$
y_N =y_1 +\ln N \ .\eqno(6)
$$
With percolation just one cluster is formed and $N$ becomes the number $\bar N_s$ of strings:
$$
y_{\bar N_s} = y_1 +\ln \bar N_s \ . \eqno(7)
$$

A simple model can be imagined where, at relatively low energy, there is a transfer of momentum from the valence string (leading proton) with creation of identical sea strings, followed, at very higher energy, by the mechamism of percolation with regeneration of fast strings, [12]. One obtains a flat distribution in rapidity ending at $y=y_{\bar N_s}$ (see (7)).

The consequences of percolation become very clear: a decrease of particle density at mid rapidity, see (3), and an extension of the length of the forward sea rapidity distribution from $\sim y_1$ to $\sim \ln \bar N_s$, see (7).

If we impose in the model energy conservation in $AA$ collision when there are $N_{part}$ participating nucleons,
$$
\int_0^{y_{MAX}} E(y) {dn \over dy} dy = {N_{part}\over 2}  {\sqrt{s}\over 2} \ ,\eqno(8)
$$
where $E(y) = \langle m_T \rangle \cosh y$ and $\sqrt{s} = m_b e^{y_b}$ we obtain, as $y_b \to \infty$,
$$
\int_0^{e^{\Delta_0}}  {\langle m_T \rangle \over m_b} {2 \over N_{part}} {dn \over d\Delta} d (e^{-\Delta}) = 1 \ , \eqno(9)
$$
where
$$
\Delta_0 =\ln \bar N_s - y_b \ . \eqno(10)
$$

If we write for the asymptotic behaviour of the number of strings
$$
\bar N_s \sim s^{\lambda} \sim e^{\lambda y_b} \ . \eqno(11)
$$
we obtain
$$
\Delta_0 =-\alpha y_b \ , \eqno(12)
$$
with
$$
\alpha = 1-2\lambda \ .  \eqno(13)
$$
One should notice that $1+\lambda$ is, in this approach, the intercept of the bare Pomeron, [12]. As $const. \leq \bar N_s \leq \sqrt{s}$ one also sees that $0\leq \alpha \leq 1$.

In order to make practical use of (9), we need to estimate the asymptotical behaviour of the ratio $\langle m_T\rangle /m_b$. We take two simple models:

\underline{Model I} - One assumes that the relation (4) for $\langle p_T\rangle$ applies as well to $\langle m_T\rangle$, [12]. This means that large $p_T$ physics is dominating everywhere.

\underline{Model II} - One assumes that the ratio $\langle m_T\rangle/m_b$ is energy independent. Such approximation is probably more realistic, as one is mostly considering the forward rapidity contributions where the $p_T$ behaviour is expected to be almost energy independent [13].

Note that in $\langle m_T\rangle$ it is also implicit an averaging over rapidity. For a discussion of this problem, see [14].

We can now estimate the asymptotic behaviour of the integrand of (9),
$$
{\langle m_T\rangle \over m_b} F(\eta) \bar N_s \bar n_1 \ \infinito e^{-\tilde \Delta _0} \ , \eqno(14)
$$
where $\tilde \Delta_0 =- \tilde \alpha y_b$ with, for central collisions, i.e., $F(\eta) \to 1/\eta^{1/2}$,
$$
\tilde \alpha = 3/2 \lambda \qquad ({\rm Model\ I}),  \eqno(15)
$$
and
$$
\tilde \alpha = \lambda \qquad \ ({\rm Model\ II}).  \eqno(16)
$$

It is clear, see Fig. 1a, that conservation of energy requires
$$
-\tilde \Delta_0 + \Delta_0 = 0 \eqno (17)
$$
or $\tilde \alpha = \alpha$. One further obtains
$$
\lambda = 2/7 \qquad ({\rm Model\ I}),  \eqno(18)
$$

$$
\lambda =1/3 \qquad \ ({\rm Model\ II}).  \eqno(19)
$$

\begin{figure}[t]
\begin{center}
\includegraphics[width=11cm]{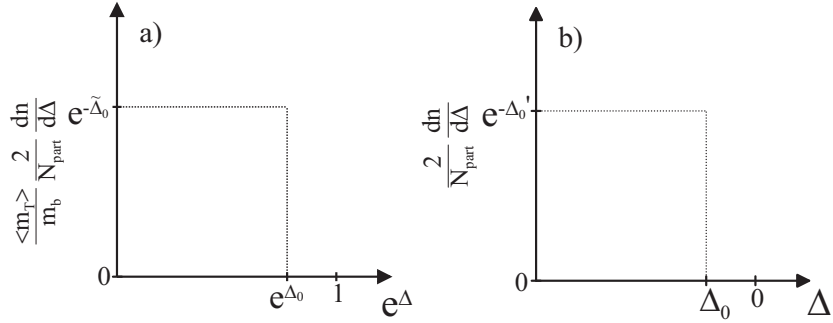}      
\end{center}
\caption{a) The integrand of (9) as a function of $e^{\Delta}$. b) The normalized $dn/d\Delta$ distribution as a function of $\Delta$. In both cases $\Delta_0 = -\alpha y_b$.}
\end{figure}

In order to arrive at the asymptotic behaviour of $dn/dy$ we simply have to divide the integrand of (9) by $\langle m_T\rangle/m_b$ (see Fig. 1b) to obtain
$$
dn/dy \sim e^{-\Delta'_0} \sim e^{\alpha' y_b} \ , \eqno(20)
$$
with 
$$
\alpha' = \alpha -1/2 \lambda = 2/7 \qquad ({\rm Model\ I}),  \eqno(21)
$$
and
$$
\alpha' = \alpha = 1/3 \qquad \ ({\rm Model\ II}).  \eqno(22)
$$

The model (with two variations) that we have considered corresponds to a step function distribution: $dn/d\Delta \equiv e^{-\Delta'_0}, \Delta \leq \Delta_0$; $dn/d\Delta =0$, $\Delta > \Delta_0$ (see Fig. 1b). It trivially satisfies limiting fragmentation, (1), with $f(\Delta)\equiv 0$.

We shall now generalize the model by introducing the Fermi distribution,
$$
2/N_{part} dn/d\Delta = {e^{-\Delta'_0} \over e^{\Delta -\Delta_0 \over \delta} + 1} \ , \eqno(23)
$$
with $\Delta'_0 = - \alpha' y_b \ , \ \Delta_0 = (1-2\lambda )y_b =-\alpha y_b$ and $\delta$ being a parameter. In the limit $\delta \to 0$ we, of course, recover the step function. In what follows we shall not distinguish between pseudo-rapidity and rapidity.

\begin{figure}[t]
\begin{center}
\includegraphics[width=10cm]{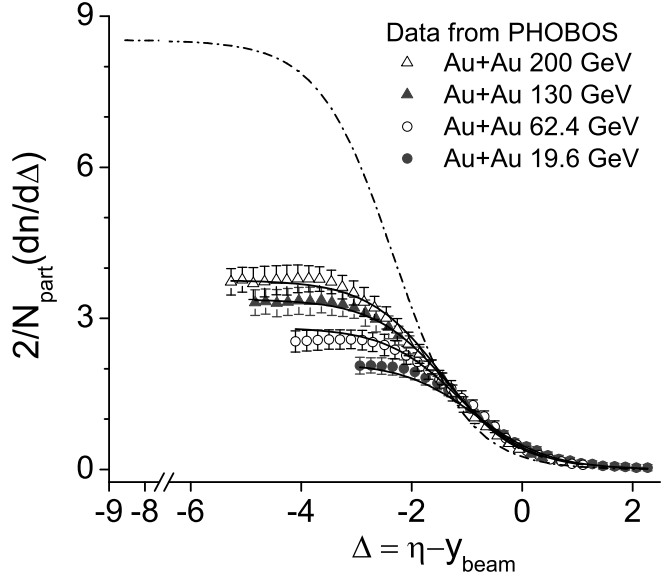}      
\end{center}
\caption{Overall fits to PHOBOS/RHIC data using Eq. (23) -- see the Table. The dotted-dashed line is a prediction for LHC, $\sqrt{s} =5500$GeV.}
\end{figure}

In the Table we present the values of $\alpha' , \alpha$ and $\delta$ in the case of Model I and Model II. We show as well the results for $\alpha' , \alpha$ and $\delta$ when fitting RHIC data for ${2\over N_{part}} dn/dy$, with Eq. (23). Our results are very similar to the fits of [15]. We note that the experimental values for $\alpha'$ and $\alpha$ are not very different, as in Model II. This means that the energy dependence of $\langle m_T\rangle/m_b$ is weak.
 
In the Table we have also included the values of $\alpha' , \alpha$ and $\delta$ resulting from an overall fit to RHIC/PHOBOS data. 

In Fig. 2 we show our curves from the overall fit, in comparison with data, and our prediction for LHC, 5500 GeV data. At mid rapidity we expect $dn/dy$ to be about 1500.

We come now back to the question of limiting fragmentation. It is clear, from Fig. 2, that we do not have strict limiting fragmentation in the $\Delta \maiorsim 0$ region.

From (23) one sees that, for $\Delta \gg \Delta_0$, limiting fragmentation simply means
$$
{2\over N_{part}} {dn\over d\Delta} \sim e^{-\Delta/\delta} \ , \eqno(24)
$$
which requires, for $\delta >0$, the limiting fragmentation condition (see (23))
$$
-\Delta'_0 + {\Delta_0\over \delta} = 0 \ , \eqno(25)
$$
or
$$ 
\delta = \Delta_0/\Delta'_0 = \alpha/\alpha' \ . \eqno(26)
$$

The fact that we do not have limiting fragmentation is not our choice. The RHIC/PHOBOS data, fitted with the parameterization (23), clearly shows that relation (26) is not obeyed: $\alpha /\alpha' \maiorsim 1$ and $\delta \simeq 2/3 < 1$ (see the Table).

In our estimates, we have assumed that at present energies and for large number of participating nucleons, $F(\eta) \to 1/\eta^{1/2} , \eta$ being large enough and increasing with energy and $N_{part}$, [10]. So we do expect changes in our parameters in (23) when moving from central (Au-Au, 0-6\%) to peripherical (Au-Au, 35-45\%) collisions. Our parameterization for peripherical collisions gives $\alpha '=0.228\pm 0.002 , \alpha = 0.235\pm 0.008$ and  $\delta = 0.90\pm 0.03$, to be compared with the values in the Table, for central collisions. At much higher energy we expect $\alpha ', \alpha$ and $\delta$ to approach the central collision values and to become the same for all centralities, as well as for $pp$ collisions.

The parameter $\Delta_0 =-\alpha y_b$ plays, in our model, the important role of controlling the separation (sharply in Models I and II) of the mid rapidity dense central region -- where percolation dominates -- from the fragmentation region where, in general, the participating nucleons retain their individuality. The central rapidity region $\Delta_{central} =-\alpha y_b -(-y_b) = (1-\alpha) y_b$ grows with the energy. But the fragmentation region, $\Delta_{fragm.} =0-(-\alpha y_b)=\alpha y_b$ also increases with the energy. This is not the case of an extended plateau (Feynmann-Wilson plateau), followed by a fixed rapidity length fragmentation region.

We finally note that the experimental value found for $\lambda$, not very different from the values of Model I and Model II, $0.36$, is consistent with values found for the intercept of the Pomeron in colour glass saturation models extended to AA scattering via geometrical scaling, [16].
\bigskip
\bigskip

{\it Acknowledgments}

\bigskip

We would like to thank J.G. Milhano and Y. Shabelski for discussions. This work has been done under contract POCTI/FNU/50210/2003 (Portugal), and FPA 2005-0196 (Spain).

\bigskip
\bigskip

\break\newpage

\centerline{Table}
\

The parameters $\alpha', \alpha$ and $\delta$ of Eq. (23), in the case of Models I and II, and as obtained from PHOBOS data, [2].

\

\begin{tabular}{c|c|c|c|}
                      & $\alpha'$       & $\alpha$      & $\delta$        \\\hline
{ Model I}            &  2/7             &  3/7          & 0               \\\hline
{ Model II}           &  1/3             &  1/3          & 0               \\\hline
{ RHIC data [2] }     &                  &               & 	               \\\hline    
$\sqrt{s} =19.6$ GeV       & $0.26  \pm 0.002$ & $0.27\pm 0.03$  & $0.65\pm 0.05$ \\
$\sqrt{s} =62.4$ GeV       & $0.230 \pm 0.008$ & $0.23\pm 0.02$  & $0.62\pm 0.07$ \\
$\sqrt{s} =130$ GeV        & $0.249 \pm 0.006$ & $0.28\pm 0.02$  & $0.67\pm 0.07$ \\
$\sqrt{s} =200$ GeV        & $0.251 \pm 0.005$ & $0.29\pm 0.01$  & $0.70\pm 0.06$ \\\hline 
{Overall fit to RHIC data} & $0.247 \pm 0.003$ & $0.269\pm 0.007$& $0.67\pm 0.03$\\\hline  
\end{tabular}

\bigskip

\bigskip

{\it References}
\bigskip

\begin{enumerate}
\item J. Benecke, J.T. Chou, C.N. Yang and E. Yen, Phys. Rev. 188, 2159 (1969).
\item B.B. Back et al. [PHOBOS Collaboration], Nucl. Phys. A757, 28 (2005); G.I. Veres, nucl-ex/0511037 (2005).
\item J. Jalian-Marian, Phys. Rev. C70, 027902 (2004).
\item B.B. Back, nucl-ex/0411012 (2004).
\item A. Capella, U.P. Sukhatme, C.I. Tan, J. Tr\^an Thanh Van, Phys. Rep. 236, 225 (1994).
\item B.Z. Kopeliovich and A. Zacharov, Z. Physik C43, 241 (1988); B.Z. Kopeliovich and B. Vogh, Phys. Lett. B446, 321 (1999).
\item I.G. Bearden et al. [BRAMS Collaboration], Phys. Rev. Lett. 93, 102301 (2004).
\item B.Z. Kopeliovich, N.N. Nikolaev and I.K. Potashnikova, Phys. Rev. D39, 769 (1989); A. Capella, C. Pajares and A.V. Ramalho, Nucl. Phys. B231, 75 (1984); J. Ranft, Phys. Rev. D51, 64 (1995); A.B. Kaidalov, K.A. Ter Materosian and Yu.M. Shabelski, Yad. Fiz. 43, 1282 (1986); N.N. Kalmikov and S.S. Ostaochanko, Phys. At. Nucl. 56, 346 (1993); K. Werner, Phys. Reports 232, 87 (1993); L. Durand and H. Pi, Phys. Rev. Lett. 58, 303 (1987); T.K. Gaisser and F. Halzen, Phys. Rev. Lett. 54, 174 (1987); H.J. Dresher, astro-ph/0411143.
\item N.S. Amelin, M.A. Braun and C. Pajares, Phys. Lett. B306 (1993); N. Armesto, M.A. Braun, E.G. Ferreiro and C. Pajares, Phys. Rev. Lett. 77, 3736 (1996); M. Nardi and H. Satz, Phys. Lett. B442, 14 (1998).
\item J. Dias de Deus and R. Ugoccioni, Phys. Lett. B491, 253 (2000); Phys. Lett. B494, 53 (2000).
\item M.A. Braun, E.G. Ferreiro, F. del Moral and C. Pajares, Eur. Phys. J. C25, 249 (2000).
\item J. Dias de Deus, M.C. Esp\'{i}rito-Santo, M. Pimenta and C. Pajares, hep-ph/0507227, to appear in Phys. Rev. Lett. (2006). 
\item J. Dias de Deus and C. Pajares, ``Percolation of colour sources and critical temperature", submitted for publication in Phys. Lett. B (2006).
\item B.B. Back, nucl-ex/0508018 (2005).
\item J. Adams et al. [STAR Collaboration], nucl-ex/0511026.
\item J.L. Albacete, N. Armesto, J.G. Milhano, C.A. Salgado and U.A. Wiedemann, Phys. Rev. D71 (2005) 014003; N. Armesto, C.A. Salgado and U. Wiedemann, Phys. Rev. Lett. 94 (2005) 022002.
\end{enumerate}
\end{document}